\documentclass[%
 reprint,
 amsmath,amssymb,
 aps,
]{revtex4-1}

\usepackage{comment}
\usepackage{physics}

\newcommand{\xj}{\mathcal{X}_{j}}

\newcommand{\br}{\mathbf{r}}
\newcommand{\brhat}{\hat{\mathbf{r}}}

\newcommand{\dint}{\textrm{d}}

\newcommand{\CX}{\mathcal{X}}

\newcommand{\CA}{\mathcal{A}}

\newcommand{\sothree}{\ensuremath{\mathcal{SO}\left(3\right)}}

\newcommand{\tj}[6]{ \begin{pmatrix}
   #1 & #2 & #3 \\
   #4 & #5 & #6 
\end{pmatrix}}

\def\CA{{\mathcal{A}}}

\usepackage{xcolor}

\usepackage{graphicx}%
\usepackage{dcolumn}%
\usepackage{bm}%

\begin{document}

\preprint{APS/123-QED}

\title{Atomic-scale representation and statistical learning of tensorial properties}

\author{Andrea Grisafi}
\affiliation{Laboratory of Computational Science and Modeling, IMX, \'Ecole Polytechnique F\'ed\'erale de Lausanne, 1015 Lausanne, Switzerland}
\author{David M. Wilkins}%
\affiliation{Laboratory of Computational Science and Modeling, IMX, \'Ecole Polytechnique F\'ed\'erale de Lausanne, 1015 Lausanne, Switzerland}
\author{Michael J. Willatt}%
\affiliation{Laboratory of Computational Science and Modeling, IMX, \'Ecole Polytechnique F\'ed\'erale de Lausanne, 1015 Lausanne, Switzerland}
\author{Michele Ceriotti}%
\affiliation{Laboratory of Computational Science and Modeling, IMX, \'Ecole Polytechnique F\'ed\'erale de Lausanne, 1015 Lausanne, Switzerland}

\begin{abstract}
This chapter discusses the importance of incorporating three-dimensional symmetries in the context of statistical learning models geared towards the interpolation of the tensorial properties of atomic-scale structures. 
We focus on  Gaussian process regression, and in particular on  the construction of structural representations, and the associated kernel functions, that are endowed with the geometric covariance properties compatible with those of the learning targets.
We summarize the general formulation of such a  symmetry-adapted Gaussian process regression model, and how it can be implemented based on a scheme that generalizes the popular smooth overlap of atomic positions representation. 
We give examples of the performance of this framework when learning the polarizability and the ground-state electron density of a molecule.
\end{abstract}

\pacs{Valid PACS appear here}%

\maketitle

\section{Introduction}

\begin{figure}[h!]
    \centering
    \includegraphics[width=4cm]{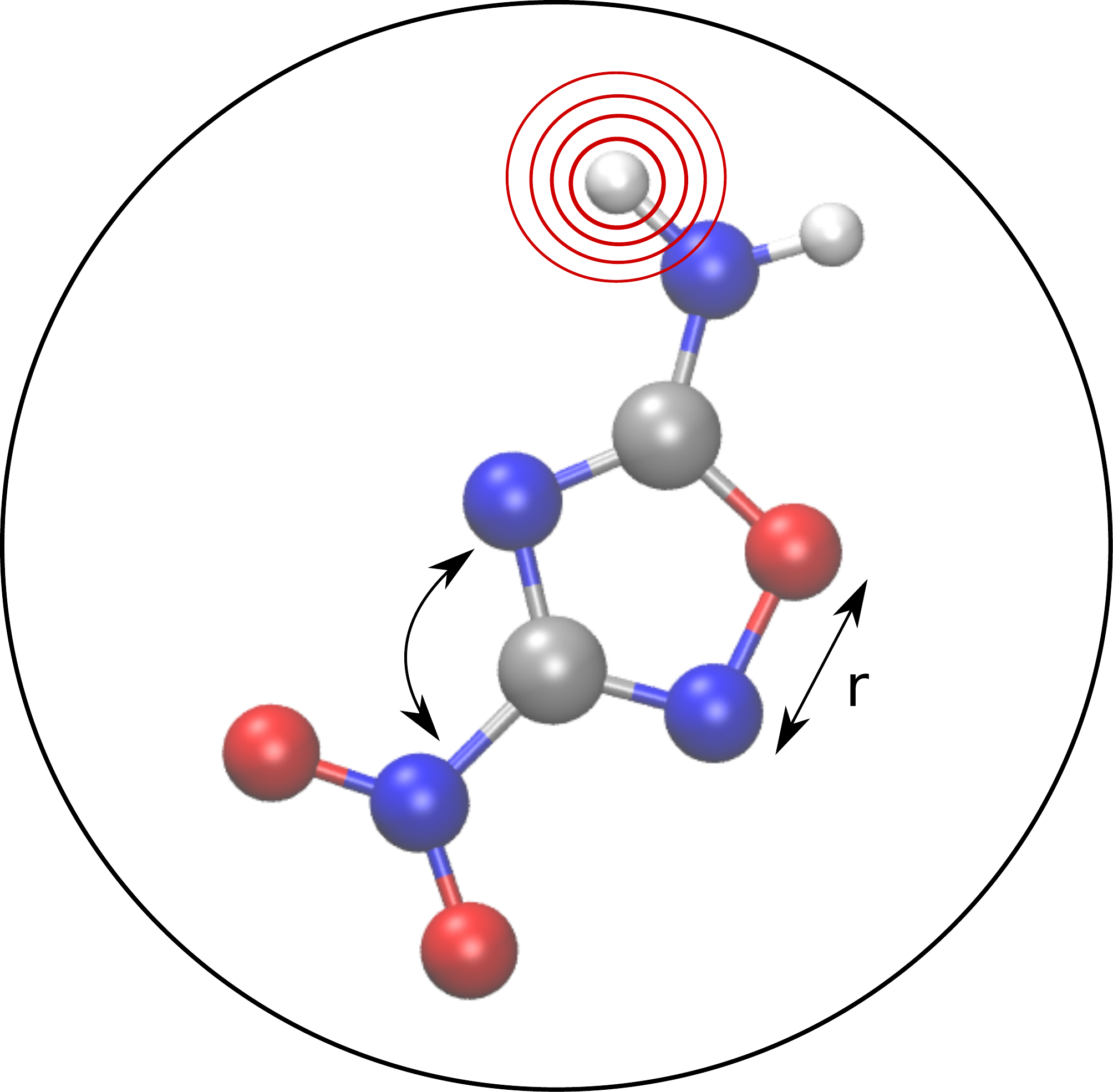}
    \caption{Structural descriptors should identify unequivocally and concisely the geometry and composition of a molecule or condensed phase.}\label{fig:representation-cartoon}
\end{figure}

The purpose of a statistical learning model is the prediction of regression targets by means of simple and easily accessible input parameters~\cite{mlbook}. In chemistry, physics and materials science, regression targets are usually scalars or tensors, including electronic energies~\cite{bartok2010,jain2013,calderon2015,ward2017}, quantum-mechanical forces~\cite{Li,Glielmo2017,Glielmo2018}, electronic multipoles~\cite{Yuan2014,Bereau2015,Bereau2017}, response functions~\cite{liang2017,grisafi2018,Wilkins2019,Christensen2019} and scalar fields like the electron density~\cite{burke2017,Alred2018,grisafi2019}. For ground-state properties, the regression input usually consists of all the information connected with the atomic structure at a given point of the Born-Oppenheimer surface, e.g. nuclear charges and atomic positions. A more or less complex manipulation of these primitive inputs leads to what is usually called a \emph{structural descriptor}, or \emph{representation} (Fig.~\ref{fig:representation-cartoon}). 

It is widely recognized that an essential ingredient for maximizing the efficiency of machine learning models is to use representations that mirror the properties one wants to predict.
Here we discuss an effective approach to build linear regression models for tensors. The notion that the representation should mirror the property means when a symmetry operation is applied to an atomic structure, the associated representation should transform in a way that mimics the transformation of the properties of the structure. It should be stressed that it is completely possible to build a machine learning model that does not incorporate such transformation properties. The universal symmetries of the property must then be learned by the model through exposure to data in the training set, making the training process less efficient. A crucial focus of this chapter is the creation of \textit{symmetry-adapted} representations.

Once one has a symmetry-adapted representation at hand, the linear regression model is bound to fulfill the symmetry requirements imposed by the property~\cite{braa-bowm09irpc,behler2007,bartok2013,Shapeev2015,Zhang2018}.  There is, however, another important consideration when building a model for tensors, expressed in terms of a Cartesian reference system. It is well known that any tensor can be decomposed into a set of spherical components that transform independently under rotations~\cite{Weinert1980,Stone}. Particularly for high-order tensors, the irreducible spherical decomposition of a tensor simplifies greatly the learning task, compared to the Cartesian representation, as we will discuss later on. 

The process of symmetry-adapting a representation is general but rather abstract, and for it to be practical one must choose the initial representation with care. For this purpose we use the Smooth Overlap of Atomic Positions (SOAP) framework, which is based on the representation of atom-centered environments constructed from a smooth atom density built up using Gaussians centered on each neighbor of the central atom. 
This density-based representation can be adapted to incorporate correlations between atoms to any order. It has been applied successfully to a vast number of machine learning investigations for physical properties of atomic structures~\cite{Sandip2016,musi+18cs,bartok2017}. 
After summarizing the derivation and efficient implementation of an extension to SOAP, called $\lambda$-SOAP, which is particularly well-suited to the learning of tensorial properties,
we present a few examples to
demonstrate its effectiveness for this task.

\section{Linear Regression}

Suppose one wanted to build a linear regression model to predict a scalar property $y(\CX)$ for an input $\CX$,
\begin{equation}
   y(\CX) = \bra{w}\ket{\CX}.
   \label{eq:linpred}
\end{equation}
In this equation $\ket{w}$ represents the weight vector we wish to learn and $\ket{\CX}$ is a representation of the input. The usual approach for learning the weight vector is to suppose the properties are independently and normally distributed, i.e.
\begin{equation}
    y(\CX) \sim \mathcal{N}(\bra{w}\ket{\CX}, \sigma^{2}).
\end{equation}
One then maximizes the log likelihood of a set of $N$ observations $\{y_{n}\}$ with respect to the weight vector. The log likelihood (loss or cost function) is given by 
\begin{equation}
    L(w) = \sum_{n=1}^{N} \sigma^{-2} (y_{n} - \bra{w}\ket{\CX_{n}})^{2} + \alpha^{2} 
    \bra{w}\ket{w},
\end{equation}
where the regularizer $\alpha^{2}\bra{w}\ket{w}$ appears if one introduces a Gaussian prior on $w$ with variance $\alpha^{-2}$. $L(w)$ attains its maximum at
\begin{equation}
    \ket{w} = \hat{C}^{-1} \sum_{n=1}^{N} \ket{\CX_{n}} y_{n},
\end{equation}
where the covariance $\hat{C}$ is
\begin{equation}
\hat{C} = \sum_{n=1}^{N} \ket{\CX_{n}}\bra{\CX_{n}} + \eta^{2} \hat{I},
\end{equation}
and $\eta = \alpha / \sigma$.

The preceding linear regression scheme in which one handles the representation $\ket{\CX}$ explicitly is often called the \textit{primal} formulation. There is in fact another, complementary formulation called the \textit{dual} (Kernel Ridge Regression (KRR) or Gaussian Process Regression (GPR)) in which the equations take a slightly different form. In the dual, one does not handle the representation explicitly but rather introduces a kernel function which -- roughly speaking -- measures the similarity between two inputs. The link between the primal and dual lies in the observation that a positive-definite kernel $k(\CX, \CX')$ can always be written as an inner product~\cite{mlbook},
\begin{equation}
   k(\CX, \CX') = \bra{\CX}\ket{\CX'}.
   \label{eq:linrepresenter}
\end{equation}
This means that given a kernel one can always construct a representation and vice versa. From the perspective of GPR, the kernel is interpreted as the covariance between the properties of its two arguments,
\begin{equation}
    \textrm{Cov}[y(\CX), y(\CX')] = k(\CX, \CX') + \sigma^{2} \delta_{\CX\CX'}.
\end{equation}
The properties are assumed to be normally distributed, which means one can straightforwardly find the conditional distribution of the property $y(\CX)$ given a set of observations in a training set $\{y_{n}\}$. The mean of this distribution is given by
\begin{equation}\label{GPR}
    y(\CX) = \mathbf{k}(\CX)^{T} \left[\boldsymbol{K} + \sigma^{2}\boldsymbol{I}\right]^{-1} \mathbf{y} = \mathbf{k}(\CX)^{T} \mathbf{x}(\{\mathbf{y}\}),
\end{equation}
where the $j$\textsuperscript{th} component of $\mathbf{k}(\CX)$ is $k(\CX, \CX_{j})$, $K_{jk} = k(\CX_{j}, \CX_{k})$ and $\mathbf{y}$ is a vector formed from $\{y_{n}\}$.

When the feature space associated with a kernel is known explicitly, and finite-dimensional, the primal and dual formulations are formally equivalent, and the choice of which to use is an important but purely practical question. Constructing a primal model requires inversion of the covariance matrix, while the dual requires inversion of the kernel matrix $\boldsymbol{K}$. If the feature space, i.e. the space occupied by the representation, is larger than the training set then the GPR approach is more convenient.  Of course, the real utility of the kernel trick becomes apparent when the kernel is a complex, non-linear function for which the feature space is unknown and/or infinite-dimensional.  In these circumstances, working in the dual makes it possible to formulate regression as a linear problem, where reference configurations (or a sparse set of representative states) is used to define a basis for the target, as in the r.h.s. of Eq.~\eqref{GPR}.
As such, all the complexity of the input space representation is contained in the definition of the kernel function.

\section{Tensors, symmetries and correlations}

The previous discussion defines the general architecture of  regression models which can be used to predict any scalar quantity associated with the molecular geometry.
We now discuss the implications of learning tensors, or, similarly, any quantity that is not invariant under a rigid rotation or reflection of the atomic structure. In so doing, we will introduce a  formalism which is general enough to encompass both proper Cartesian tensors, such as molecular polarizabilities, and three-dimensional scalar fields that can be conveniently decomposed in atom-centered contributions, such as the ground-state charge density of a molecule.

Let us start by considering the prototypical case of a Cartesian tensor $y\equiv y_{\alpha\beta...}$ of rank $r$, with the combination of indices $\{\alpha\beta...\}$ running over a number of Cartesian components equal to $3^r$. Given any arbitrary distorted atomic structure with no particular internal symmetry, we are interested in characterizing the transformations of the tensor under only three families of symmetry operations, viz. \emph{translations}, \emph{rotations} and \emph{reflections}. Since these symmetry operations do not affect the internal geometry of an atomic structure, we can think equivalently in terms of active transformations, in which the system undergoes the symmetry operation and the reference frame remains fixed, or in terms of passive transformations, in which the reference frame undergoes the symmetry operation and the system remains fixed. In the following, we summarize the symmetry operations by adopting an active picture and assume the system is not subjected to an external field.

\emph{Translations.} Any physical property of an atomic structure $\CX$ remains unchanged under a rigid translation $\hat{t}$ of atomic positions, i.e.,
\begin{equation}
    y_{\alpha\beta...}(\hat{t}\CX) = y_{\alpha\beta...}(\CX)\ .
    \label{translations}
\end{equation}
~
 \emph{Rotations.} Under the application of a rigid rotation $\hat{R}$ to an atomic structure $\CX$, we assume that each Cartesian component of the tensor undergoes a covariant linear transformation. Using Einstein notation for convenience, and representing by $\boldsymbol{R}$ the rotation matrix corresponding to $\hat{R}$, the rotated tensor is \begin{equation}\label{rotations}
    y_{\alpha\beta...}(\hat{R}\CX) = R_{\alpha\alpha'} R_{\beta\beta'}\times ... \times y_{\alpha'\beta'...}(\CX)\ .
\end{equation}

\emph{Reflections.} Applying a reflection operator $\hat{Q}$ to an atomic structure $\CX$ through any mirror plane leads to the following reflected tensor,
\begin{equation}\label{reflections}
    y_{\alpha\beta...}(\hat{Q}\CX) = Q_{\alpha\alpha'} Q_{\beta\beta'}\times ... \times y_{\alpha'\beta'...}(\CX)\ .
\end{equation}

\subsection{Covariant descriptors}

In general terms, a primitive representation that mirrors a tensor of a given rank $r$ could formally be built by considering
\begin{equation}\label{primitive_descriptor}
\ket{\CX \alpha\beta...}  =  \ket{\CX} \otimes  \ket{\alpha}  \otimes \ket{\beta}  \otimes ...,    
\end{equation}
where $\ket{\CX}$ is an arbitrary description of the system, while~$\ket{\alpha}$ represents a set of Cartesian axes which is rigidly attached to the system. 
When using this primitive representation in a linear regression model, the tensor component corresponding to $\alpha \beta...$ would be
\begin{equation}
    y_{\alpha\beta...}(\CX) = \bra{w}\ket{\CX\alpha\beta...},
\end{equation}
or
\begin{equation}
y_{\alpha\beta...}(\CX) = \bra{w_{\alpha\beta...} }\ket{\CX\alpha\beta...}.
\end{equation}
After maximizing the log likelihood, the former possibility leads to a model that predicts every component to be the same, while the latter ignores the known correlations between the components and is therefore likely to overfit. For example, consider a training set in which only one of the tensor components is non-zero. All but one of the regression weights $\{\ket{w_{\alpha\beta...}}\}$ would be driven towards zero to maximize the log likelihood, so the trained model would only predict a finite value for the component it had been explicitly exposed to in the training set. The model would therefore incorrectly predict the tensor components for a structure differing only by a rigid rotation from one in the training set. 

\begin{figure}
    \centering
    \includegraphics[width=8cm]{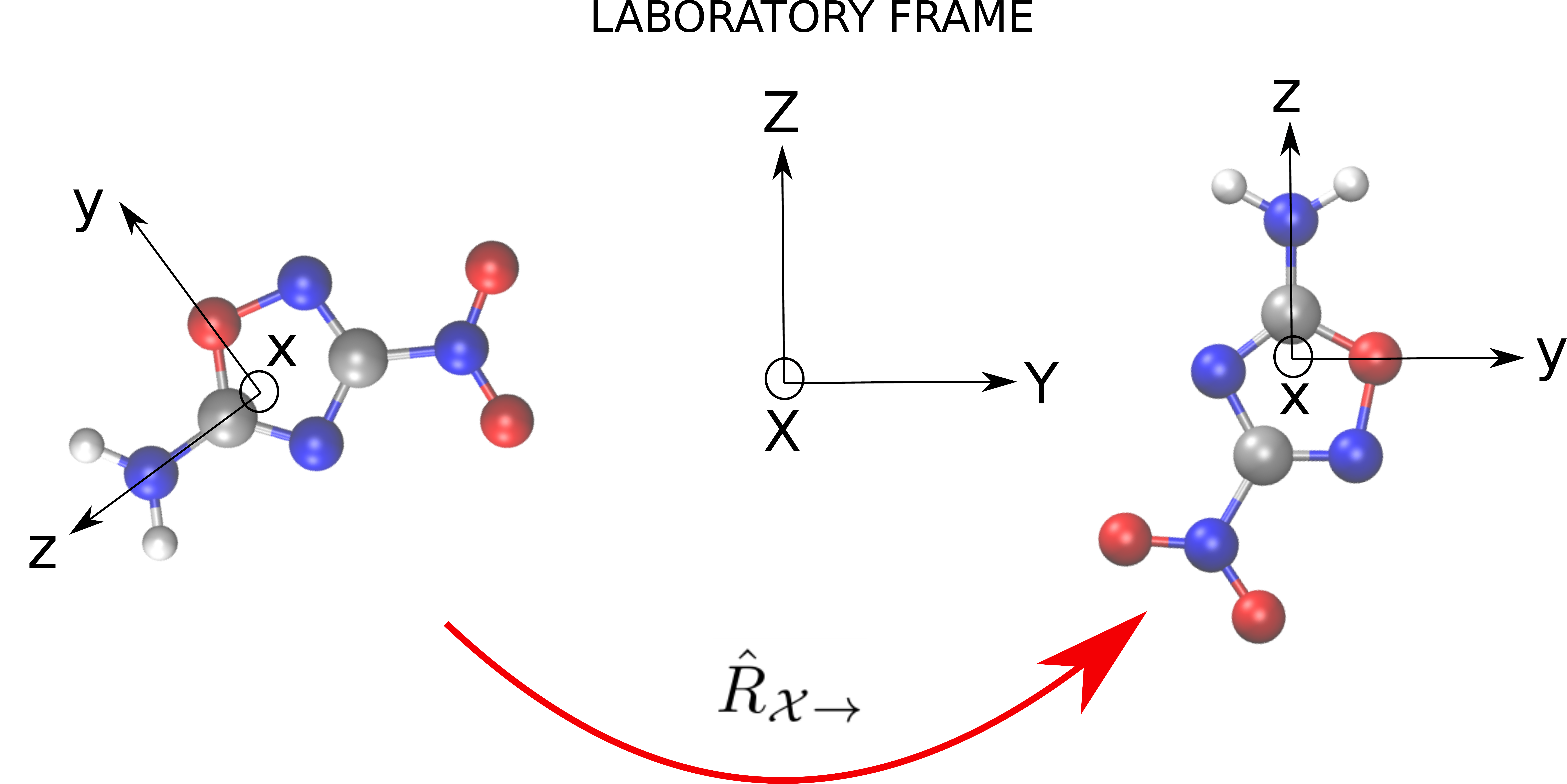}
\caption{Provided that one can define a local reference system, it is possible to learn tensorial properties by aligning each molecule (or environment) into a fixed reference frame.}
    \label{fig:ext_alignment}
\end{figure}

To address these problems, one should adapt the primitive descriptor so that it fulfills each of the symmetries detailed in Eqs.~(\ref{translations}-\ref{reflections}). 
Since the Cartesian basis vectors are invariant under translations, Eq.~(\ref{translations}) implies the core representation should itself be invariant under translations. Using Haar integration one can construct a core representation that is invariant under translations by integrating an arbitrary representation over the translation operator $\hat{t}$~\cite{willatt2018arxiv}.
One can then proceed to consider covariance under $\sothree$ group operations. Eq.~(\ref{rotations}) implies that a covariant representation for $\ket{\CX\alpha\beta...}$ should satisfy the invariance relationship
\begin{equation}
    \label{eq:rotinvar}
  \left[\hat{I} \otimes \hat{R} \otimes \hat{R} \otimes ...\right] \ket{
  \left(\hat{R}\CX\right)\alpha\beta...}_{\sothree} =  \ket{\CX\alpha\beta...}_{\sothree},
\end{equation}
for any rotation $\hat{R}$. Starting from the primitive definition of Eq.~\eqref{primitive_descriptor}, there are a variety of ways to enforce this invariance relationship. One possibility is to use
\begin{equation}\label{aligned descriptor}
\ket{\CX\alpha\beta...}_{\sothree} \equiv 
 \left[\hat{R}_{\CX\to} \otimes \hat{R}_{\CX\to} \otimes ...\right] \ket{\left(\hat{R}_{\CX\to}\CX\right)\alpha\beta...}
\end{equation}
where the operator $\hat{R}_{\CX\to}$ is defined to rotate $\CX$ into a specified orientation which is common to all the molecules of the dataset (Fig.~\ref{fig:ext_alignment}). 

This works under the assumption that it is always possible to define a unique (and therefore unambiguous) internal reference frame to rotate $\CX$ into a specified orientation, which might be possible when the system involved has a particularly rigid internal structure. 
A more general strategy, which does not require any assumption on the molecular geometry to be made, consists in considering the covariant integration over the operator $\hat{R}$ (\emph{Haar integration}),
\begin{equation}
\label{eq:haarrotation}
\ket{\CX\alpha\beta...}_{\sothree} \equiv \int  \dint\hat{R}
 \left[\hat{I} \otimes \hat{R} \otimes ...\right] \ket{\left(\hat{R}\CX\right)\alpha\beta...}.
\end{equation}
On the top of this definition, the requirement that a representation be covariant in $\mathcal{O}(3)$, including the reflection symmetry of the tensor as in Eq.~\eqref{reflections}, means that a simple linear combination of the $\sothree$ descriptor with its reflected counterpart with respect to any arbitrary mirror plane of the system must be considered; that is,
\begin{equation}\label{o3_descriptor}
    \ket{\CX\alpha\beta...}_{\mathcal{O}(3)} =  \ket{\CX\alpha\beta...}_{\sothree} + \left[\hat{I} \otimes \hat{Q}\otimes ...\right] \ket{\left(\hat{Q}\CX\right)\alpha\beta...}_{\sothree} 
\end{equation}
for any arbitrary choice of $\hat{Q}$. Any other reflection operation can be automatically included by having made the descriptor covariant under rotations.

\subsection{Covariant regression}

Having shown how to build a symmetry-adapted  representation of the system, let us see the implications of this procedure for linear regression. Using a symmetry-adapted representation in a linear regression model leads to the following solution for the regression weight,
\begin{equation}
    \ket{w} = \sum_{n=1}^{N} \sum_{\alpha\beta...} \hat{C}^{-1} \ket{\CX_{n}\alpha\beta...}_{\mathcal{O}(3)} y_{\alpha\beta...}(n),
\end{equation}
where the covariance is
\begin{equation}
    \hat{C} = \sum_{n=1}^{N} \sum_{\alpha\beta...} \ket{\CX_{n}\alpha\beta...}_{\mathcal{O}(3)}\bra{\CX_{n}\alpha\beta...}_{\mathcal{O}(3)} + \eta \hat{I}.
\end{equation}
Note that the solution for the linear regression weight does not change when the training structures and corresponding tensors simultaneously undergo a symmetry operation that the representation has been adapted to. In other words, the same model results regardless of the arbitrary orientation of structures in the training set.

When moving to the dual, we find the kernel to be
\begin{equation}
    k_{\alpha\beta...}^{\alpha'\beta'...}(\CX, \CX') = 
    \bra{\CX'\alpha'\beta'...}\ket{\CX\alpha\beta...}_{\mathcal{O}(3)}.
\end{equation}
This result corresponds to 
\begin{equation}
    \int  \dint \hat{R} \int  \dint R' \bra{\hat{R}\CX}\ket{\hat{R}'\CX'} (RR')_{\alpha \alpha'} (RR')_{\beta \beta'}...
\end{equation}
As stressed earlier, performing the linear regression in the dual using this kernel leads to a formally-equivalent model to that resulting from the primal formulation described above, yet this kernel appears to be more complicated than a symmetry-adapted descriptor since it involves two integrations over rotations. If, however, we assume the core representation $\ket{\CX}$ undergoes a unitary transformation when the system is rotated, 
\begin{equation}
    \ket{\hat{R}\CX} = \hat{U}[\hat{R}] \ket{\CX},
\end{equation}
the kernel reduces to
\begin{equation}\label{covariant_kernel}
     k_{\alpha\beta...}^{\alpha'\beta'...}(\CX, \CX') = 
    \int  \dint \hat{R} \, k(\CX, \hat{R}\CX') R_{\alpha \alpha'} R_{\beta \beta'}...,
\end{equation}
where $k(\CX, \CX') = \bra{\CX}\ket{\CX'}$ is the kernel corresponding to the core representation. The requirement that the core representation should undergo a unitary transformation when the system is rotated is reasonable since, if it were not true, the autocorrelation $k(\CX, \CX)$ would depend on the absolute orientation of $\CX$, which is unphysical given our assumption of the absence of external fields.

Note that upon defining a collective tensorial index $\{\alpha\beta...\}$, a kernel matrix of size $3^r N\times3^r N$ can be constructed by stacking together each of the $3^r\times3^r$ vector-valued correlation functions~\cite{Alvarez2012}. Then, a covariant tensorial prediction of the property of interest can eventually be carried out according to the GPR prescription of Eq.~\eqref{GPR}.

\begin{figure}
    \centering
    \includegraphics[width=9cm]{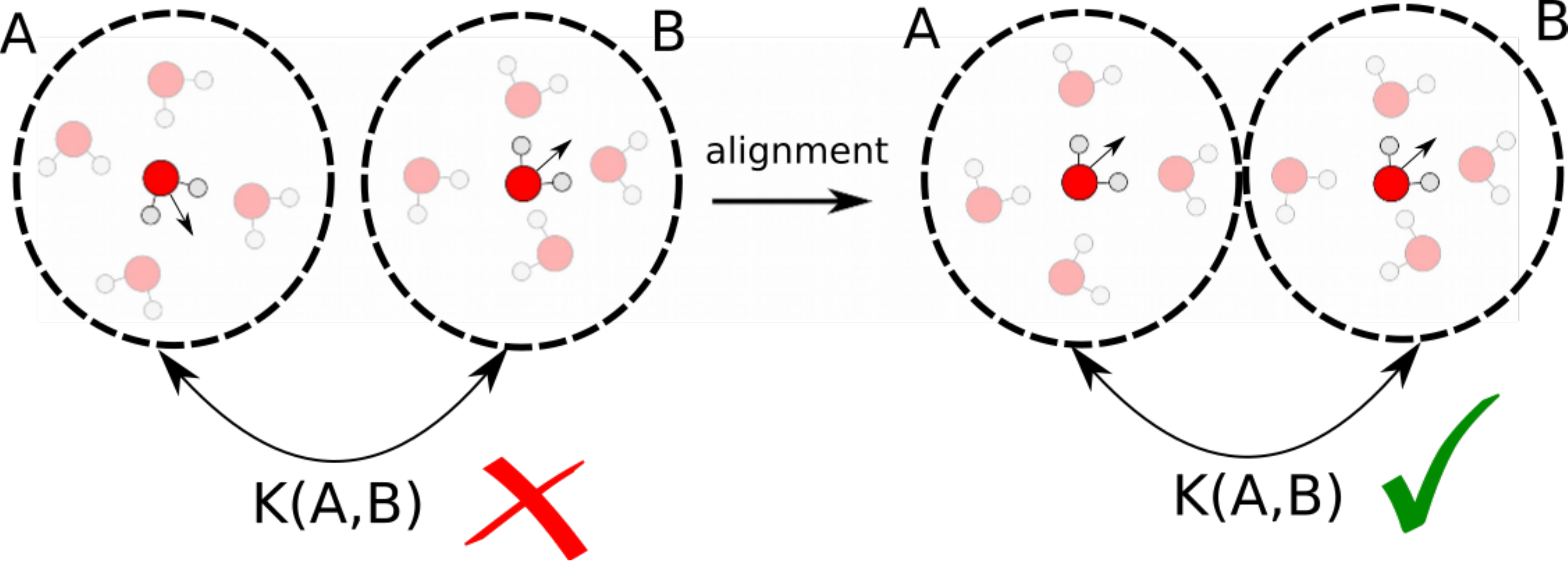}
    \caption{Representation of the reciprocal alignment between water environments.}
    \label{fig:alignment}
\end{figure}

It is instructive to compare the symmetry-adapted kernel definition of Eq.~\eqref{covariant_kernel} to the kernel that one gets from the aligned descriptors of Eq.~\eqref{aligned descriptor}. In this case, building a kernel function on the top of this descriptor effectively means carrying out the structural comparison in a common reference frame where the two molecules are mutually aligned. One can then conveniently learn the tensor of interest component-by-component through a much simpler scalar regression framework. For the simple case of rank-1 tensors, for instance, we would get,
\begin{equation}\label{alignment_kernel}
   \boldsymbol{k}_{\sothree}(\CX,\CX') =  \langle\CX'|\hat{R}_{\CX'\to\CX}|\CX\rangle  \boldsymbol{R}_{\CX'\to\CX},
\end{equation}
where we have defined the best alignment operator as $\hat{R}_{\CX'\to\CX}=\hat{R}_{\CX'\to}\hat{R}^T_{\CX\to}$.
This strategy has been successfully used in the learning of electronic multipoles of organic molecules~\cite{Bereau2015} as well as for predicting optical response functions of water molecules in their liquid environments~\cite{liang2017}. For the latter example, a representation of the best-alignment structural comparison is reported in Fig.~\ref{fig:alignment}.

This method for tensor learning has the clear drawback of relying on the definition of a rigid molecular geometry, for which an internal reference frame can be effectively used to perform the procedure of best alignment.  Following this line of thought, the availability of a covariant kernel function allows us to implicitly carry out both the structural comparison and the geometric alignment of two molecules simultaneously, neglecting any prior consideration about the internal structure of the molecule at hand.

\subsection{Spherical representation}

The family of Cartesian symmetry-adapted descriptors previously introduced can be effectively used, in principle, to predict any Cartesian tensor of arbitrary order. However, we should notice that having a tensor product for each additional Cartesian axis makes the cost of the regression scale unfavorably with the tensor order, producing a global kernel matrix of dimension $(3^r)^2$. 
In fact, it is well established that a more natural representation of Cartesian tensors is given by their irreducible spherical components (ISC)~\cite{Stone}.
As described in Ref.~\cite{Stone}, the transformation matrix from Cartesian to spherical tensors can be found recursively, starting from the known transformation for rank-2 tensors.

Upon trivial manipulations, the components of a spherical tensor transform separately as the irreducible representations of the $\sothree$ group. Each $\lambda$-component of the tensor spans an orthogonal subspace of dimension $2\lambda+1$. For instance, the 9 components of a rank-2 tensor separate out into a term (proportional to the trace) that transforms like a scalar, three terms that transform like $Y^1_m$, and five terms that transform like $Y^2_m$.  When using a spherical representation, the kernel matrix is block diagonal, which greatly reduces the number of non-zero entries, and makes it possible to learn separately the different components.  An additional advantage is that the possible symmetry of the tensor can be naturally incorporated by retaining only the spherical components $\lambda$ that have the same parity as the tensor rank $r$. For instance, the $\lambda=1$ component of a symmetric rank-2 tensor vanishes identically, meaning that only the 6 surviving elements of the tensor need to be considered when doing the regression. Especially for high rank tensors, this property means that the number of components can be cut down significantly.

In light of the discussion carried out for Cartesian tensors, it is straightforward to realize how a symmetry-adapted descriptor that transforms covariantly with spherical harmonics of order $\lambda$ should look.  Since each ISC is effectively a vector of dimension $2\lambda+1$, we can first write a primitive spherical harmonic representation as
\begin{equation}\label{primitive_spherical_descriptor}
\ket{\CX\lambda\mu} = \ket{\CX} \otimes \ket{\lambda\mu},\
\end{equation}
where $\ket{\lambda\mu}$ is an angular momentum state of order $\lambda$, such that $\bra{\hat{\br}}\ket{\lambda\mu}=Y^\lambda_\mu(\hat{\br})$.
Its symmetry-adapted counterpart, which is covariant in $\sothree$, is
\begin{equation}\label{spherical_descriptor}
\ket{\CX\lambda\mu}_{\sothree} = \int  \dint\hat{R}\ \ket{\hat{R}\CX} \otimes \hat{R}\ket{\lambda\mu}\ .
\end{equation}
Finally, since odd spherical harmonics are anti-symmetric with respect to the inversion operator $\hat{i}$, a spherical tensor descriptor that is covariant in $\mathcal{O}(3)$ can be obtained by considering
\begin{equation}\label{inversion_symmetry}
\ket{\CX\lambda\mu}_{\mathcal{O}(3)} =  \ket{\CX\lambda\mu}_{\sothree} + (-1)^\lambda\ \ket{\left(\hat{i}\CX\right)\lambda\mu}_{\sothree}\ .
\end{equation}
Note that a tensorial kernel function built on the top of this descriptor would transform under rotations as the Wigner-$D$ matrix of order $\lambda$, $D_{\mu\mu'}^\lambda=\langle\lambda\mu|\hat{R}|\lambda\mu'\rangle$:
\begin{equation}
\boldsymbol{k}^\lambda_{\sothree}(\mathcal{X},\mathcal{X}') = \int  \dint\hat{R}\ \bra{\CX}\ket{\hat{R}\CX'} \boldsymbol{D}^\lambda(\hat{R})
\end{equation}

In addition to being the most natural strategy to perform the regression of Cartesian tensors, using a representation like that of Eq.~\eqref{inversion_symmetry} comes in handy when building regression models for the many physical properties that can be decomposed in a basis of atom-centered spherical harmonics. In the following sections, we will give an example of this kind by predicting the ground-state electronic charge density of molecular systems.

\section{\label{sec:lambdasoap} SOAP representation}

We now proceed to characterize the exact functional form of a symmetry-adapted representation of order $\lambda$ which can be used to carry out a covariant prediction of any property that transforms as a spherical harmonic. In Sec. II it was pointed out that, within a framework of linear regression, both the primal and the dual formulation can be adopted to actually implement the interpolation of a given tensorial property. 
In what follows, however, we will focus our attention on the dual formulation, discussing in parallel the feature vector associated with the $\lambda$-SOAP representation and the corresponding kernel function.  This choice is justified by the greater flexibility of the kernel formulation, allowing a non-linear extension of the framework as discussed below.

An atom-centered environment $\CX_j$ describes the set of atoms that are included within a spherical cutoff $r_\text{cut}$ around the central atom $j$. We will label as $\ket{\CX_j}$ the abstract vector which describes the local structure.
A convenient definition of $\ket{\CX_j}$ in real space can be obtained by writing a smooth probability amplitude, for each atomic species $\alpha$, as a superposition of Gaussians with spread $\sigma$ that are centered on the positions $\{\br_i\}$ of the atoms that surround the central atom $j$:
\begin{equation}
    \psi^\alpha_{\CX_j}(\br) = \sum_{i \in \CX_j, \alpha} \exp\left\{-\frac{|\br-\left(\br_i-\br_j\right)|^2}{2\sigma^2}\right\}
\end{equation}
This definition descends naturally from the requirement of translational invariance of a representation of the entire structure~\cite{willatt2018arxiv} and corresponds to the construction that is used in Ref.~\cite{bartok2013} to define the Smooth Overlap of Atomic Positions (SOAP) kernel.
Formally, one can then write
\begin{equation}
\bra{\br}\ket{\CX_j}=\sum_\alpha \ket{\alpha}\psi^\alpha_{\CX_j}(\br)
\end{equation}
with the ket $\ket{\alpha}$ tagging the identity of each species. Even though it might be convenient to use a lower-dimensional chemical space~\cite{will+18pccp}, particularly when building models for dataset containing many elements, in what follows we will assume that each element is associated with an orthogonal subspace, i.e. $\bra{\alpha}\ket{\beta}=\delta_{\alpha\beta}$. This implies that, when using this representation to define a scalar-product kernel,  only the density distributions of the same atomic type are overlapped, 
\begin{equation}
\bra{\CX_{j'}}\ket{\CX_j}=
\sum_\alpha \int \dint\br\, \psi^\alpha_{\CX_{j'}}(\br) \psi^\alpha_{\CX_j}(\br)
\end{equation}
With this choice, the two adjustable parameters $r_\text{cut}$ and $\sigma$ determine respectively the range and the resolution of the representation.
To simplify the notation, we will omit the $\alpha$ labels, assuming that a single element is present. The extension to the case with multiple chemical species follows straightforwardly.

\subsection{$\lambda$-SOAP(1) representation}

To the first order in structural correlations, including the environmental state $\ket{\CX_j}$ in the definition of a local symmetry-adapted descriptor of order $\lambda$ reads 
\begin{equation}\label{lambda_soap_1}
   \ket{\mathcal X^{(1)}_{j}\lambda\mu}_{\sothree} =  \int  \dint\hat{R}\ \hat{R}\ket{\CX_j} \otimes \hat{R}\ket{\lambda\mu}.
\end{equation}
The real space representation of $\ket{\mathcal X^{(1)}_{j}\lambda\mu}$ can be understood as a rotational average of the environmental density  which is rigidly attached to a spherical harmonic of order~$\lambda$,
\begin{equation}
\label{eq:tresp1}
   \bra{\mathbf{r}\hat{\mathbf{r}}' }\ket{\mathcal{X}^{(1)}\lambda \mu}_{SO(3)} = \int \dint \hat{R} \, \mathcal{X}_{j}(\hat{R} \mathbf{r}) Y^\lambda_\mu(\hat{R} \hat{\mathbf{r}'}).
\end{equation}

A more concise, and easily-computed version of this representation results from projecting $\ket{\mathcal X^{(1)}_{j}\lambda\mu}_{\sothree}$ on a basis of spherical harmonics, in which the integral over rotations can be performed analytically,
\begin{equation}
\begin{split}
    & \bra{rlml'm'}\ket{\mathcal{X}^{(1)}_{j}\lambda \mu}_{SO(3)} 
    \propto \frac{r}{2\lambda + 1} \delta_{ll'} \delta_{\lambda l'}
    \delta_{mm'}
    \bra{r\lambda\mu}\ket{\mathcal{X}_{j}}. 
\end{split}
\end{equation}
It is clear that many of the indices in this representation are redundant, and would have no effect when taking an inner product between two such representations. The most concise form that produces the same scalar product kernel as Eq.~\eqref{eq:tresp1} is 
\begin{equation}
   \bra{r}\ket{\mathcal{X}_{j}^{(1)}\lambda\mu}_{SO(3)} \equiv \frac{r}{\sqrt{2\lambda + 1}} \bra{r\lambda\mu}\ket{\mathcal{X}_{j}},\label{eq:lsoap1-mini}
\end{equation}
where we introduced the spherical density component 
\begin{equation}
\bra{rlm}\ket{\CX_{j}} = \int\dint \brhat\  Y^{l}_m(\brhat)^\star\psi_{\CX_j}(r\brhat).
\end{equation}
This ket corresponds to the kernel
\begin{equation}
   k^{\lambda}_{\mu\mu'}(\mathcal{X}_{j}, \mathcal{X}_{k}) = \frac{1}{2\lambda + 1} \int \dint r\, r^2 \bra{\mathcal{X}_{j}}\ket{r\lambda\mu}\bra{r\lambda\mu}\ket{\mathcal{X}_{k}},
\end{equation}
which is straightforward to calculate using e.g. quadrature in $r$ or an expansion on a radial basis.

It is insightful to consider the explicit expression for Eq.~\eqref{eq:lsoap1-mini} in terms of the atom density. Taking for instance the case of $\lambda=1, \mu=0$, for which $Y^1_0(\brhat)\propto x/r$: 
\begin{equation}
\bra{r}\ket{\mathcal{X}_{j}^{(1)}10}_{SO(3)} \propto
\int\dint \br\, x \psi_{\CX_j}(\br) \delta(r-|\br|).
\end{equation}
One sees that the 2-body $\lambda$-SOAP representation corresponds to moments of the smooth atom density, resolved over different shells around the central atom. 
A linear model built on these features can respond to changes in the atomic density at different distances, simultaneously adapting the magnitude and geometric orientation of the target property.

\subsection{$\lambda$-SOAP(2) representation}

Describing an atomic environment in a way that goes beyond the two-body  structural correlations ($\nu>1$) is of fundamental importance, because information on distances alone is not sufficient to uniquely determine an atomic structure.  Building on the definition of Eq.~\eqref{lambda_soap_1}, and on the symmetrized-atom-density framework of Ref.~\cite{willatt2018arxiv}, this can be achieved by introducing an additional tensor product in the environmental state $\ket{\CX_j}$ within the rotational average,
\begin{equation}\label{lambda_soap_2}
   \ket{\xj^{(2)}\lambda\mu}_{\sothree} =  \int  \dint\hat{R}\ \hat{R}\ket{\CX_j} \otimes \hat{R}\ket{\CX_j} \otimes \hat{R}\ket{\lambda\mu}\ .
\end{equation}
By projecting on a real-space basis, the representation becomes
\begin{equation}
\label{eq:resp1}
    \bra{\mathbf{r}\mathbf{r}'\hat{\mathbf{r}}''}\ket{\mathcal{X}^{(2)}_{j}\lambda \mu}_{SO(3)} = \int dR \, \mathcal{X}_{j}(R \mathbf{r}) \mathcal{X}_{j}(R\mathbf{r}') Y_{\lambda \mu}(R \hat{\mathbf{r}}'').
\end{equation}

Similarly to the $\nu=1$ case, one can compute the ket without an explicit rotational average by projecting on a basis of spherical harmonics,
\begin{equation}
\begin{split}
    & \bra{rr'lml'm'l''m''}\ket{\mathcal{X}^{(2)}_{j}\lambda \mu}_{SO(3)} 
    \propto \delta_{\lambda l''} \tj{l}{l'}{\lambda}{m}{m'}{m''} \\
    & \times rr' \sum_{kk'} \tj{l}{l'}{\lambda}{k}{k'}{\mu}
    \bra{rlk}\ket{\mathcal{X}_{j}} \bra{r'l'k'}\ket{\mathcal{X}_{j}},
\end{split}
\end{equation}
where the parentheses denote a Wigner 3j symbol. Just as for the $\lambda$-SOAP(1) case considered earlier, it is clear that many of the indices in this expression are redundant. When taking an inner product between two such representations, one can use orthogonality of Wigner 3j symbols to simplify to an inner product between two objects with the following form,
\begin{equation}
 \bra{rr'll'}\ket{\mathcal{X}^{(2)}_{j}\lambda \mu}_{SO(3)} 
    \equiv rr'  
     \sum_{kk'} \bra{lk,lk'}\ket{\lambda\mu}
    \bra{rlk}\ket{\mathcal{X}_{j}} \bra{r'l'k'}\ket{\mathcal{X}_{j}}.
\end{equation}
The Clebsch-Gordan coefficient $\bra{lk,lk'}\ket{\lambda\mu}$ has the role of combining two angular momentum components of the atomic environment $\mathcal{X}_{j}$ to be compatible with the spherical tensor order $\lambda$. This contains all the essential information of the abstract representation $\ket{\mathcal{X}^{(2)}_{j}\lambda \mu}_{SO(3)}$ that is needed for $\lambda$-SOAP(2) linear regression. Note that $\bra{lk,lk'}\ket{\lambda\mu}$ is zero unless $k+k'=\mu$,  that the indices $l$, $l'$ and $\lambda$ must satisfy the inequality $|l - l'| \le \lambda \le l + l'$ and that the representation is invariant under transposition of $r$ and~$r'$.

Let us see how the representation changes under inversion. Given the parity of the spherical harmonics,
\begin{equation}
    \bra{rlk}\ket{\hat{i}\mathcal{X}_{j}} = (-1)^{l} \bra{rlk}\ket{\mathcal{X}_{j}},
\end{equation}
it follows that
\begin{equation}
\begin{split}
    & \bra{rr'll'}\ket{\hat{i} \mathcal{X}^{(2)}_{j}\hat{i}\lambda \mu}_{SO(3)}  = (-1)^{l + l' + \lambda} \bra{rr'll'}\ket{\mathcal{X}^{(2)}_{j}\lambda \mu}_{SO(3)}.
\end{split}
\end{equation}
This condition implies that a representation that is covariant in $\mathcal{O}(3)$ can be easily obtained by retaining only the components of the feature vectors for which  $l + l' + \lambda$ is even. 
Generalization of this procedure to higher orders of $\lambda$-SOAP is tedious but straightforward using well-known formulae for integrals of products of Wigner-$D$ matrices over rotations.

\subsection{Non-linearity}

As already mentioned in the introduction, a crucial aspect to improve regression performance is to incorporate non-linearities in the construction of the representation. For instance, tensor products of the scalar representation introduce higher body order correlations, in a way that can be easily implemented in a kernel framework by raising the kernel to an integer power~\cite{willatt2018arxiv}. 
When working with tensorial representations, however, one has to be careful to avoid breaking the covariant transformation properties of the feature vector. 
Taking products of $\ket{\CX_{j}^{(\nu)}\lambda\mu}_{\mathcal{O}(3)}$ kets would require re-projecting the product onto the irreducible representations of the group, which would be as cumbersome as increasing the body order exponent $\nu$. 
One obvious solution to this problem is to multiply the spherical kernel of order $\lambda$ by its scalar and rotationally invariant counterpart, which can then be raised to an integer power $\zeta$ without breaking the tensorial nature of the kernel. For any generic order $\nu$ and $\nu'$ in structural correlations, this procedure consists in considering the tensor product
\begin{equation}
\ket{\CX_{j}^{(\nu)}\lambda\mu}_{\mathcal{O}(3)}
 \otimes \prod^{\zeta-1}\ket{\CX_{j}^{(\nu')}00}_{\mathcal{O}(3)},
\end{equation}
which leads to the kernel definition
\begin{equation}
    \boldsymbol{k}^\lambda_{\zeta}(\CX_j,\CX_{j'}) =\boldsymbol{k}^\lambda(\CX_j,\CX_{j'}) \left(k^0(\CX_j,\CX_{j'})\right)^{\zeta-1}.
\end{equation}
For $\zeta=1$, one recovers the original tensorial kernel, while a non-linear behavior is introduced for $\zeta>1$. A considerable improvement of the learning power is usually obtained when using $\zeta=2$, while negligible further improvement is observed for~$\zeta>2$. 

\begin{figure}
    \centering
    \includegraphics[width=8cm]{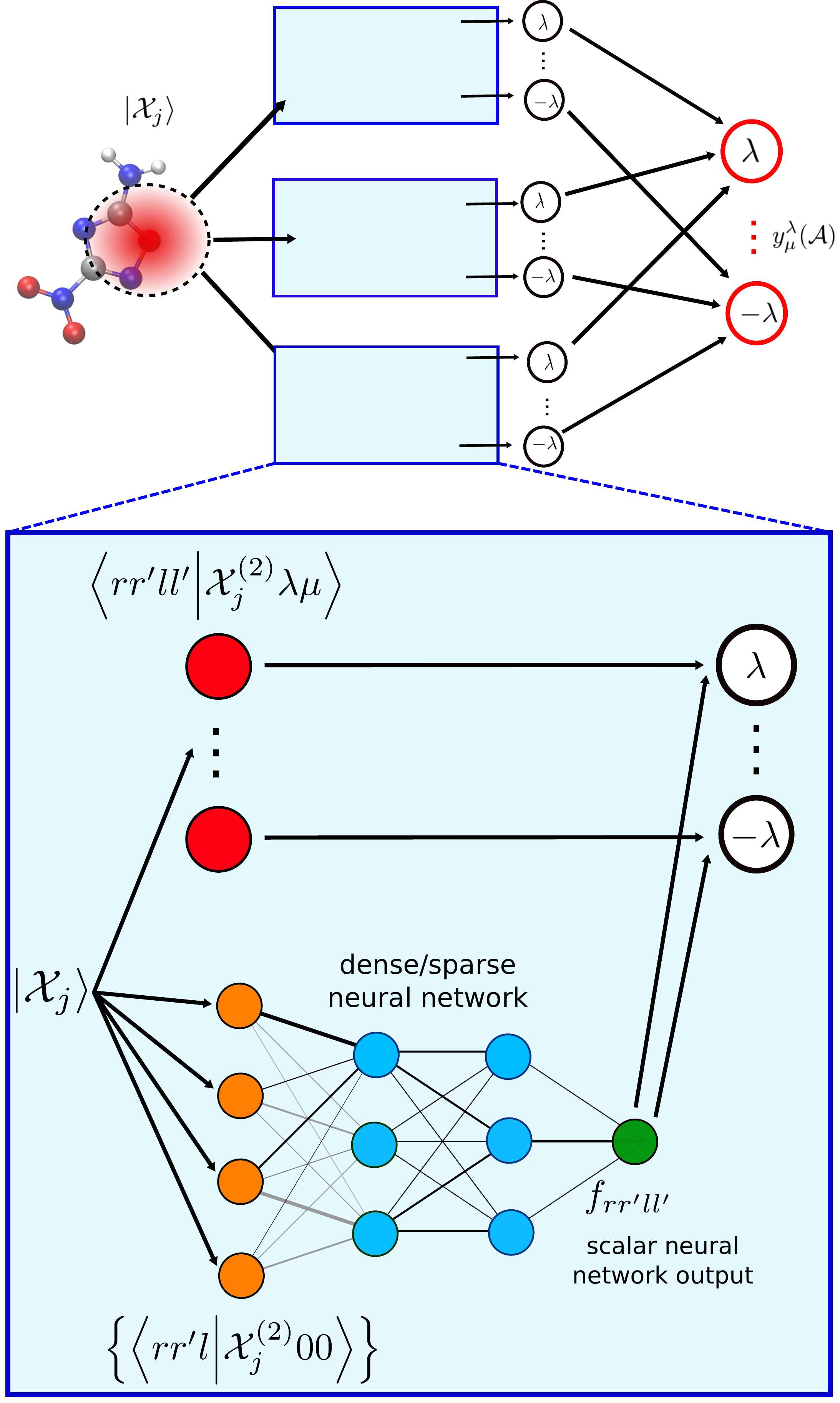}
    \caption{A schematic representation of a covariant NN architecture based on tensorial and scalar $\lambda$-SOAP representations.}
    \label{fig:covariant_NN}
\end{figure}

These considerations also apply to the use of fully non-linear machine-learning models like a neural network. To guarantee that the prediction of the model is consistent with the group covariances, the tensorial $\lambda$-SOAP features must enter the network at the last layer, and all the previous non-linear layers can only contribute to different linear combinations of the tensorial features, e.g.
\begin{equation}
y^\lambda_\mu(\CA) = \sum_{j\in\CA} 
f_{rr'll'}\left[\left\{\bra{rr'l'}\ket{\CX_j^{(2)}00}\right\}\right] \bra{rr'll'}\ket{\CX_j^{(2)}\lambda\mu},
\end{equation}
where each of the $f_{rr'll'}$ can be an arbitrary non-linear combination of the scalar SOAP features (see Fig.~\ref{fig:covariant_NN}).
Similar ideas have already been implemented in the context of generalizing the construction of spherical convolutional neural networks~\cite{kondor2018arxiv}.

\subsection{Implementation}

In the previous discussion it was pointed out that beyond the formal definition of the structural descriptor in real space, the kernel evaluation eventually requires the computation of the SOAP density power spectrum $\bra{ r r' ll'}\ket{\CX_{j}^{(2)}\lambda\mu}$. In turn, computing this quantity requires the evaluation of the density expansion coefficients~$\bra{r l m}\ket{\xj}$. In practice, the continuous variable $r$ can be replaced by an expansion over a discrete set of orthogonal radial functions $R_n(r)$ that are defined within the spherical cutoff $r_\text{cut}$. For this reason, we will refer, from now on, to the density expansion coefficients as $\bra{n l m}\ket{\xj}$.

Having represented the environmental density distribution as a superposition of Gaussian functions centered on each atom, the  spherical harmonics projection can be carried out analytically~\cite{Kaufmann}, leading to:
\begin{equation}
\begin{split}\label{semiformal_coeffs}
    \bra{n l m}\ket{\xj} = &\sum_{i} Y_{lm}(\hat{ \br}_{ij})\ \exp\left\{-\frac{|\mathbf{r}_{i}-\mathbf{r}_{j}|^2}{2\sigma^2}\right\}\ \times \\& \times \int_0^\infty \mathrm{d}r\ r^2\ R_n(r)
\exp\left\{-\frac{r^2}{2\sigma^2}\right\} \iota_l\left(\frac{r r_{ij}}{\sigma^2}\right), 
\end{split}
\end{equation}
where the sum over $i$ runs over the neighboring atoms of a given chemical element, and $\iota_l$ represents a modified spherical Bessel function of the first kind. Under suitable choices of the functions $R_{n}(r)$, the radial integration can also be carried out analytically, too.

One possibility is to start with non-orthogonal Gaussian type functions, $\tilde{R}_k(r)$, reminiscent of Gaussian-type orbitals commonly used in quantum chemistry:
\begin{equation}
    \tilde{R}_{k}(r) = \mathcal{N}_k\ r^{k} \exp{-\frac{1}{2}\left(\frac{r}{\sigma_{k}}\right)^2}, 
\end{equation}
where $\mathcal{N}_k$ is a normalization factor, such that $\int_0^\infty dr r^2 \tilde{R}^2_{k}(r) = 1$. The set of Gaussian widths $\{\sigma_k\}$ can be chosen to effectively span the radial interval involved in the environment definition. 
For instance, one can take $\sigma_k = r_\text{cut} \max(\sqrt{k},1)/n_\text{max}$, 
obtaining functions that have equally-spaced peaks between 0 and $r_\text{cut}$. The explicit functional form of the primitive radial integrals is
\begin{equation}
\begin{split}\label{primitive_coeffs}
& \int_0^\infty \mathrm{d} r\ r^2\ \tilde{R}_k(r)\
e^{-\frac{r^2}{2\sigma^2}} \iota_l\left(\frac{r r_i}{\sigma^2}\right) = \\ 
& \times \ \mathcal{N}_k\ 2^{-\frac{1}{2}(1+l+k)}\
    \left(\frac{1}{\sigma^2}+\frac{1}{\sigma_{k}^2}\right)^{-\frac{3+l+k}{2}}\ 
    \frac{\Gamma(\frac{3+l+k}{2})}{\Gamma(\frac{3}{2}+l)}  
    \\& \times\ \left(\frac{r_i}{\sigma^2}\right)^l\ {}_1F_1\left(\frac{3+l+k}{2},\frac{3}{2}+l;\ \frac{1}{2}\frac{\sigma^2_{k}\ r_i^2}{\sigma^4+\sigma_{k}^2\sigma^2}\right),
\end{split}
\end{equation}
where $\Gamma$ is the Gamma function, while ${}_1F_1$ is the confluent hypergeometric function of the first kind. These primitive integrals can be finally orthogonalized by applying the orthogonalization matrix $\boldsymbol{S}^{-1/2}$, with $\boldsymbol{S}$ representing the overlap matrix between primitive functions,
\begin{equation}
S_{kk'} = \int_0^\infty dr\ r^2 \tilde{R}_k(r)\ \tilde{R}_{k'}(r),
\end{equation}
for which well-known analytical expressions exist~\cite{Gradshteyn2007}.

\section{Examples}

In this section, the effectiveness of a kernel ridge regression model that is adapted to the fundamental physical symmetries of the target  is demonstrated, considering two very different quantities as examples.  
The first example involves the prediction of the first order dipole polarizability $\boldsymbol{\alpha}$ for a broad dataset of organic molecules, when training a $\lambda$-SOAP(2) regression model on the irreducible spherical components of the tensor $\boldsymbol{\alpha}$. In the second example, we show how to predict  the charge density $\rho(\mathbf r)$ of a small, yet flexible, hydrocarbon molecule like butane, by decomposing $\rho(\mathbf r)$ into atom-centered spherical harmonics. In both cases, a comparison of the prediction performance is carried out between $\lambda$-SOAP(2) descriptors that are covariant in $\sothree$ - which were used in previous work - and those that have been made fully $\mathcal{O}(3)$ compliant by symmetrization over $\hat{i}$.

\subsection{\label{sec:alphaml}Molecular Polarizabilities}

The polarizability of a molecule is a rank-2 tensor describing the second-order response of the molecular energy $U$ with respect to an applied electric field $\mathbf{E}$, with components $\alpha_{ij} = \partial^{2} U/\partial E_{i} \partial E_{j}$. By construction this tensor is symmetric, meaning that it can be decomposed into two components: a $\lambda=0$ (scalar) component,
\begin{subequations}
\begin{equation}
\alpha^{(0)} = -\frac{1}{\sqrt{3}} \left[ \alpha_{xx} + \alpha_{yy} + \alpha_{zz}\right],
\end{equation}
and a $\lambda=2$ (5-vector) component,
\begin{equation}
\mathbf{\alpha}^{(2)} = \sqrt{2} \left( \alpha_{xy},\alpha_{yz},\alpha_{xz},\frac{2\alpha_{zz}-\alpha_{xx}-\alpha_{yy}}{2\sqrt{3}}, \frac{\alpha_{xx}-\alpha_{yy}}{2}\right).
\end{equation}
\end{subequations}

\begin{figure}
    \centering
    \includegraphics[width=9cm]{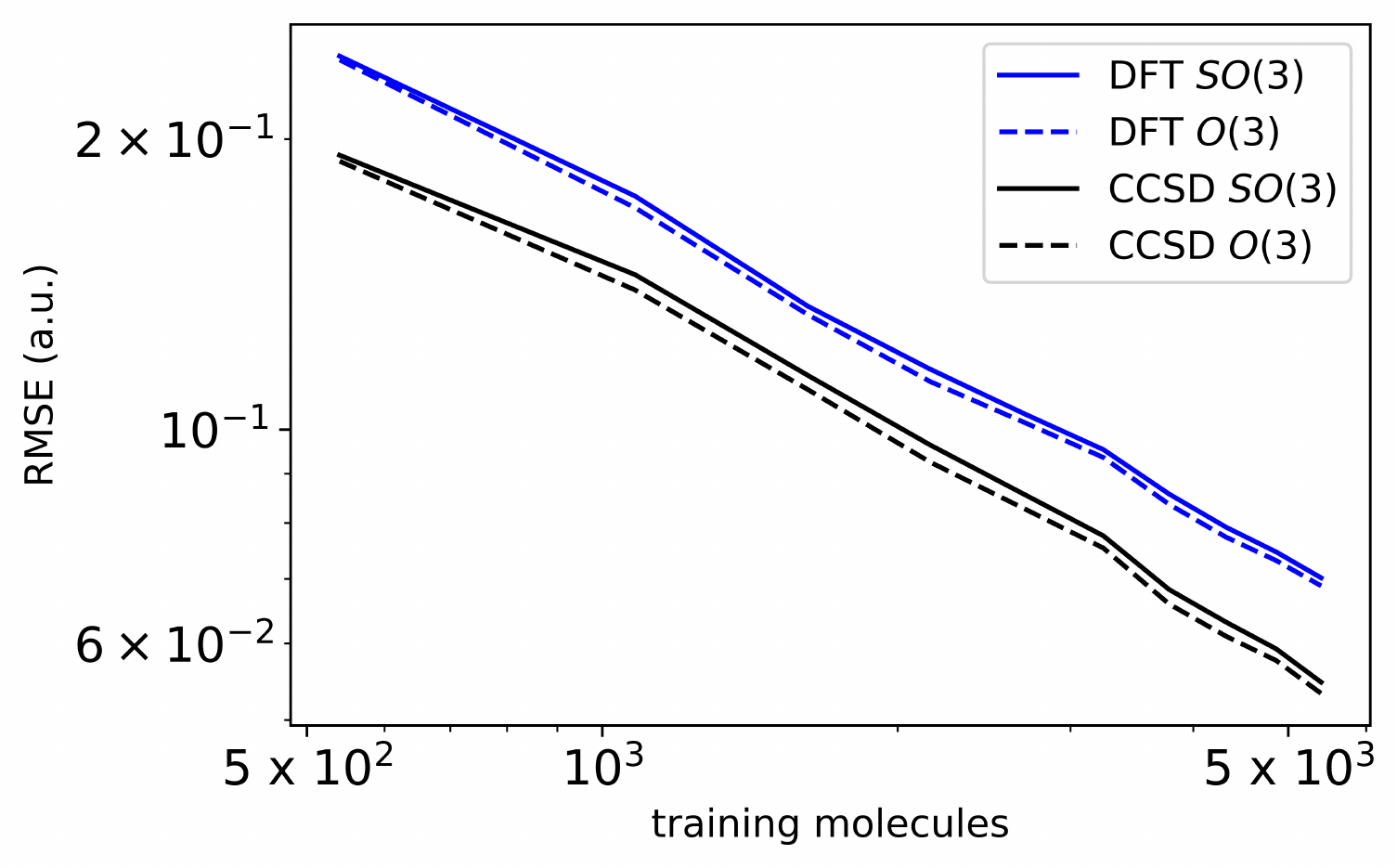}
    \caption{Learning curves of the QM7b polarizability tensors computed both at the DFT (\emph{blue lines}) and CCSD (\emph{black lines}) level. Full and dashed lines refer to predictions that are carried out with $\lambda$-SOAP kernel functions that are covariant in $\mathcal{SO}$(3), as reported in Ref.~\cite{Wilkins2019}, and $\mathcal{O}$(3) respectively. In all cases, the testing set consists of 1,811 molecules.} 
    \label{fig:SO3_vs_O3_qm7b}
\end{figure}

Fig.~\ref{fig:SO3_vs_O3_qm7b} shows learning curves for the prediction of the $\mathbf{\alpha}^{(2)}$ component for polarizabilities of molecules in the QM7b database~\cite{mont+13njp}, which contains 7,211 small molecules. The results are shown for two types of kernel: the first of these is an $\sothree$ kernel as used in Ref.~\cite{Wilkins2019} and the second of this is an $\mathcal{O}(3)$ kernel accounting for inversion symmetry. 
All kernels were built using 8 radial functions, maximum spherical harmonic cutoff $l_{\text{ max}}=6$, $\sigma=0.3~\text{\AA}$ and $\zeta=2$, and kernels with different values of the radial cutoff $r_{\text{cut}}$ were combined to form a multiscale kernel\cite{bartok2017} as $\boldsymbol{k}^{\lambda}(\mathcal{X},\mathcal{X}^\prime) = \sum_{i} c_{i} \boldsymbol{k}^{\lambda}_{i}(\mathcal{X},\mathcal{X}^\prime)$, where $\boldsymbol{k}^{\lambda}_{i}(\mathcal{X},\mathcal{X}^\prime)$ is the kernel with $r_{\text{ cut}}/\text{\AA} = i$ and $c_{2} = 0.04053$, $c_{3} = 0.00997$, $c_{4} = 0.02250$, $c_{5} = 0.01560$.
The use of a $\mathcal{O}(3)$  reduces approximately by half the size of the feature vector, and at the same time it leads to a small - but consistent - reduction in the test error. 

\subsection{Electronic charge densities}

\begin{figure}
    \centering
    \includegraphics[width=9cm]{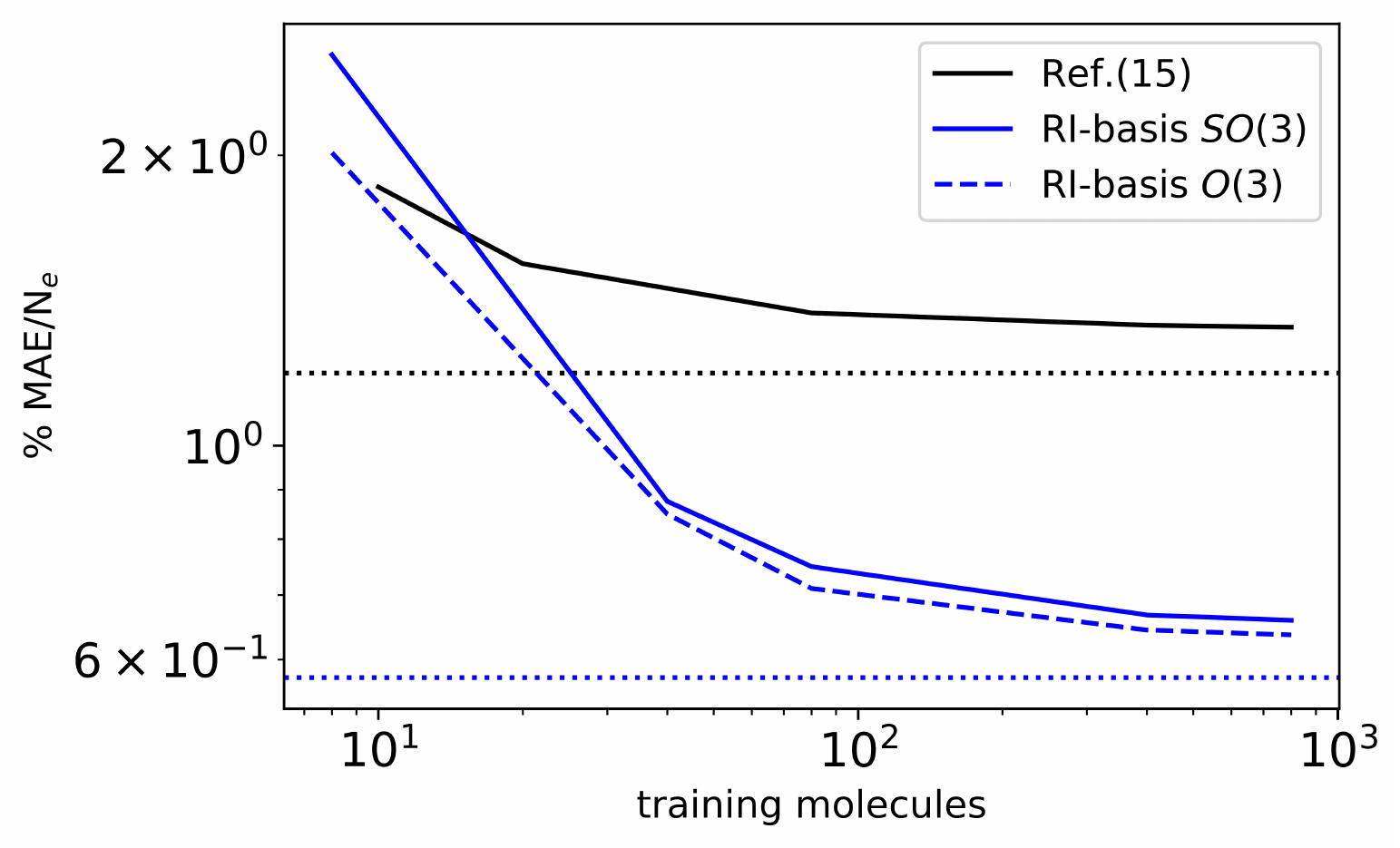}
    \caption{Learning curves of the predicted charge density of 200 randomly selected butane molecules, when considering up to 800 reference molecules to train the model. The molecular geometries and computational details are the same as in Ref.~\cite{grisafi2019}. The black full line refers to the prediction error as reported in Ref.~\cite{grisafi2019}. Blue lines refer to the result obtained with the RI-cc-pV5Z basis, both with a $\lambda$-SOAP(2) descriptor covariant in $\sothree$ (\emph{full}) and $\mathcal{O}(3)$ (\emph{dashed}). Dotted lines refer to the basis set error. In both cases, 100 reference atomic environments have been used to define the problem dimensionality.} 
    \label{fig:SO3_vs_O3}
\end{figure}

Another learning task that can benefit from a symmetry-adapted regression scheme involves the learning of scalar fields such as the electron charge density. Machine-learning models for the charge density have been proposed based on the coefficients in a plane wave basis~\cite{burke2017} -- this is convenient due to orthogonality, but leads to poor transferability when considering flexible molecules, or learning across different molecular species -- or based on direct prediction of the density on a real-space grid~\cite{Alred2018,Chandrasekaran2019}. 
By expanding the density on an atom-centred basis set, composed of radial functions multiplied by spherical harmonics,
\begin{equation}\label{charge_decomposition}
    \rho(\br) = \sum_{inlm}c^i_{nlm}\phi_{nlm}(\br-\br_i)
\end{equation}
one obtains a model that is localized and transferable, concise, and easily integrated with the many electronic structure codes that are based on atom-centered basis functions. 
The coefficients in the expansion transform under rotations like spherical harmonics, and can therefore be learned efficiently using a SA-GPR model,
\begin{equation}\label{pred_coeffs}
    c^i_{nlm} = \sum_{j\in Z_i}\sum_{|m'|<l}k^l_{mm'}(i,j)\ x^j_{nlm'}
\end{equation}
where the sum runs over a set of reference environments $Z_i$ centered around atoms of the same kind as $i$, and the weights are computed by a regression procedure that is complicated by the fact that the basis set is not orthogonal~\cite{grisafi2019}. 

In Fig.~\ref{fig:SO3_vs_O3} we report the result obtained for a dataset of butane molecules (C$_4$H$_{10}$), for which 1000 reference pseudo-valence densities have been computed at the DFT/PBE level. 
The dimensionality of the regression problem is defined by considering the 100 most diverse atomic environments, out of a total of 14000, selected by farthest point sampling\cite{ceri+13jctc} through the 0-SOAP(2) distance metric.
Given that in our previous work the learning performance was essentially limited by the basis set expansion error for the density, we decided to compare the optimized basis set used in Ref.~\citenum{grisafi2019} with a resolution of the identity (RI) basis set, usually adopted in the context of avoiding the computation of the four-center Hartree integral in electronic structure theory~\cite{Hattig2005}. When considering in particular the RI-cc-pV5Z basis, which accounts for basis functions up to $l$=4 of angular momentum, we find that the basis set decomposition error is almost halved ($\sim$0.6\%) with respect to Ref.~\citenum{grisafi2019}, as shown by the asymptotic convergence in Fig.~\ref{fig:SO3_vs_O3}.  
The figure also compares, in the case of the RI basis, 
the learning performances associated with  $\lambda$-SOAP(2) descriptors that have been made covariant in $\sothree$ and $\mathcal{O}$(3) respectively. As seen for the case of polarizability, the $\mathcal{O}$(3) features improve -- although only slightly -- the prediction accuracy. The improvement is more substantial at the smallest training set size, where the incorporation of prior knowledge on the symmetries of the system can make up for the scarcity of data.

\section{Outlook}

The previous examples show how statistical learning of a tensorial quantity across the configurational space of atomic coordinates and composition represents a challenging methodological task which requires considerable modifications to the architecture of more familiar scalar learning models. 
The efficiency of a regression model benefits greatly from the incorporation of symmetry, as it effectively reduces the dimensionality of the space in which the algorithm is asked to interpolate the values of the target property. 
Symmetry of tensorial quantities should be included in two distinct ways. First, one should decompose the tensor into irreducible spherical components, so as to minimize the amount of information that is needed to account for geometric covariance.  Particularly for high-rank Cartesian tensors, the matrix of correlations between tensor elements can be made block diagonal, which reflects on the size and complexity of the associated kernel matrices.
Second, by constructing representations of the molecular structure that are made isomorphic with the tensor of interest, one can obtain a linear basis that satisfies the expected covariant transformations. 
An important aspect to consider is that, in order to preserve the properties of the symmetry-matched basis, non-linearities have to be treated with care. We discuss how it is possible to do so in the context of kernel ridge regression models, and how one should proceed to design a covariant neural network that can be used to efficiently accomplish a symmetry-adapted regression task. 

We discuss a practical implementation of these ideas within the framework of the smooth overlap of atomic positions representations, that uses a spherical-harmonics representation of the atom density and is therefore particularly well-suited to incorporate $\sothree$ covariance.  We discuss an extension, that we refer to as $\lambda$-SOAP, that provides a natural linear basis to regress quantities that transform like spherical harmonics, and can be made to represent arbitrarily high body-order correlations between atomic coordinates. 
As an original result of this work, we also discuss how to satisfy the inversion symmetry of the tensor, showing that representations that incorporate the full $\mathcal{O}(3)$ covariances improve the performance of the machine-learning model, particularly in the limit of a small training set. 
We also show an example of the use of $\lambda$-SOAP representations to learn a scalar field in three-dimension as a sum of atom-centered contributions, choosing the electron density as a physically relevant example. We believe that this strategy -- although more complex than alternatives that use orthogonal basis functions or a real-space grid -- has the best promise to be transferable across different systems, and to be combined with standard electronic structure packages.

\begin{acknowledgments}
The Authors acknowledge support by  the European Research Council
under the European Union's Horizon 2020 research and innovation programme
(Grant Agreement No. 677013-HBMAP).
\end{acknowledgments}

\bibliographystyle{ieeetr}

\end{document}